\documentclass[preprint,showpacs,preprintnumbers,amsmath,amssymb,prb,12pt]{revtex4}
\usepackage{dcolumn}
\usepackage{graphicx,epsfig}
\usepackage{rotating,texdraw}

\usepackage{bm}

\begin{document}

\preprint{APS/123-QED}

\title{Dynamical structure factor of a nonlinear Klein Gordon lattice}

\author{Laurent Proville}
\email{lproville@cea.fr}
\affiliation{\small Service de Recherches
de M\'etallurgie Physique, CEA-Saclay/DEN/DMN
        91191-Gif-sur-Yvette Cedex, France}

\date{\today}

\begin{abstract}

The quantum modes of a nonlinear Klein Gordon lattice have been
computed numerically [L. Proville, Phys. Rev. B {\bf 71}, 104306
(2005)]. The on-site nonlinearity has been found to lead to phonon
bound states. In the present paper, we compute numerically the
dynamical structure factor so as to simulate the coherent
scattering cross-section at low temperature. The inelastic
contribution is studied as a function of the on-site
anharmonicity. Interestingly, our numerical method is not limited
to the weak anharmonicity and permits to study thoroughly the
spectra of nonlinear phonons.
\end{abstract}


\maketitle

\section{Introduction}
The metal hydrides are typical compounds whose the technological
importance\cite{Dantzer} aroused many spectroscopic inquiries. The
spectroscopy of hydrogen modes allows to work out the occupied
interstitial gaps, e.g. octahedral or tetrahedral and thus the
phase structure of the alloy. Further, the determination of the
anharmonicity of hydrogen sites\cite{Eckert,Ross1998} permits to
evaluate the on-site potential landscape\cite{Ross1998} and so the
hydrogen ability to migrate or to dimerise. The inelastic neutron
scattering (INS) revealed the on-site anharmonicity in the model
hydrides as NbH, TaH and PdH (for a review see Ref.
\onlinecite{Ross}). In the compounds TiH and ZrH, that
anharmonicity was found\cite{Kolesnikov,KolesZrH} to lead to the
optical phonon bound states, i.e., some well-known excitations in
molecular crystals\cite{bogani2,AGRANO,Califano,Gellini,Mao} (also
quoted as biphonon and bivibron) where the anharmonicity of
internal covalent bonds overpasses the interaction between
molecules. As in the metal hydrides, the proton dynamics has been
studied by INS in molecular crystals as polyglycine\cite{Fillaux0}
and 4-methyl pyridine\cite{Fillaux2}. In the latter, the bound
states of the methyl group rotational modes proved to last several
days\cite{Fillaux2,Fillaux1} (see also Refs.
\onlinecite{Wattis,ScottLastChapter}). The trapping of energy,
whether it is in proton vibrations or in the intrinsic modes of
molecules is a consequence of the emergence of breather, a
theoretical paradigm that has been matter of intensive research in
different contexts (for instance see Refs.
\onlinecite{Siev,MA94,Aubry,mackay2000,Kopidakis,Sievers2002,
Flach2002,Fleurov1998,Fleurov2003,Eilbeck,Rosenau,Tretiak,Gomez}).
The distinctive property of breather is to gather the spatial
localization and time periodicity. For the case of a
translational-invariant quantum lattice, the phonon bound states
can be thought as the siblings of
breathers\cite{Aubry,bishop96,mackay2000,LPXarch}, as phonon bound
states and breathers both stem from the lattice anharmonicity. The
formers appeared, though, earlier in literature (see Refs.
\onlinecite{Fleurov2003,Bogani} and references therein). The link
between breather and phonon bound states has been studied in
different
works\cite{ScottLastChapter,Fleurov1998,Fleurov2003,LPXarch,LPquant-phys}.
Consequently, the experiments that enhanced the existence of
phonon bound states (for instance see Refs.
\onlinecite{Kolesnikov,bogani2,Gellini}) are equally some concrete
evidences of the emergence of quantum breathers. In addition,
several experiments\cite{acetanilide,PtCl,Adachi} were recently
dedicated to breather and its important role in the energy storage
and transport. As an overview, one may retain that the studies on
breather aim at improving the standard harmonic treatment of
lattice dynamics which flaws are well-known in
solids\cite{Ashcroft2}.

The scattering of neutrons\cite{Lovesey}, X-rays\cite{Xrays} and
electrons\cite{electrons} provided numerous insights into
condensed matter and molecular physics. In particular, the
inelastic scattering allows to probe the dynamics of crystals and
molecules. The dynamical structure factor, $S(q,\omega)$ which is
proportional to the scattering cross-section can be calculated by
solving the Schr\"{o}dinger equation for the lattice modes (within
Born approximation), so this quantity is essential for bridging
theory to experiment. However the lattice eigen-modes may be
figured out exactly only in the harmonic lattice model (see for
instance Ref. \onlinecite{Lovesey,Ashcroft}) and thus the
anharmonic contributions to the energy are neglected, although
they stem from the atomic interaction potential. In order to
examin the effect of anharmonicity on $S(q,\omega)$, some
approaches have been attempted in different quantum lattice
models, e.g., the Hubbard model for
bosons\cite{AGRA1976,Dubovski1997} and the Klein Gordon model with
a weak on-site anharmonicity\cite{bishop96}. The latter is often
quoted as nonlinear Klein Gordon lattice (KG) and, noteworthy, it
may account for the intrinsic anharmonicity of molecular
crystals\cite{bogani2,Mao,Califano,Gellini}, molecule
bonds\cite{benzen,Jerzembeck} or metals where interstitial gaps
are filled with light particles, as metal
hydrides\cite{Eckert,Ross1998,Kolesnikov,KolesZrH}. It is the
reason why we proposed a numerical method\cite{LPXarch} that is
tractable for different type of nonlinearity, to compute the
nonlinear quantum modes.
In the present paper, our previous developments are used to
evaluate the coherent Dynamical Structure Factor (DSF) of the KG
lattice, at low temperature. The contribution to the DSF of the
nonlinear quantum modes, known as either the phonon bound
states\cite{AGRANO,bogani2,Kolesnikov,Califano,Gellini} or else
the quantum breathers\cite{bishop96,Flach2002,mackay2000}, is the
central purpose of our work. As the DSF standard derivation is
obtained in the harmonic approximation\cite{Lovesey,Ashcroft},
treating the anharmonicity as a perturbation, we propose a
different scheme where this approximation is not required. For
instance, the Bloch identity, which is a key for the conventional
calculation, is not invoked in our theory. We simply introduce a
Taylor expansion of the DSF with respect to the atomic
displacements, computing numerically the coefficients of the
series. Our method is tested by comparison to the standard
analytical calculation on the purely harmonic lattice. Our
approach allows us to deal with different strength of the
anharmonicity. When the on-site nonlinearity dominates the
intersite coupling, the nonlinear quantum modes lead to some
anharmonic resonances well separated from the multi-phonon
continua, in the inelastic spectrum. These anharmonic resonances
have same order as for fundamental phonons whereas the unbound
multi-phonons yield a DSF at least one order below. On the other
hand, their dispersion is found to decrease dramatically with the
value of the energy transfer, which differentiates them from the
broad phonon resonance.
By way of contrast, when the intersite coupling is larger than the
on-site nonlinearity, the anharmonicity of the spectrum is weak
compared to the phonon dispersion. In such a quasi-harmonic
lattice, the dynamical response of the biphonon (a two phonon
bound state) may yet exhibit a significant magnitude, still much
larger than the two-phonon, even though the biphonon resonance
occurs at an energy transfer that approaches the unbound phonons
band. Actually, the biphonon signature is found to dominate the
two-phonon DSF provided that the lattice nonlinearity is not
strictly zero. Alongside our numerics, a perturbation theory is
developed and proves accurate for strongly nonlinear lattices.
We also show how the model parameters may be adjusted so as to
simulate the inelastic scattering of a material and thus to work
out the energy landscape of the inner particles as well as their
interactions. Finally, the present study will serve as a basis for
future simulations of the KG incoherent scattering cross-section.

After a brief introduction to the nonlinear KG lattice, the DSF is
derived in Sec. \ref{Sec2}.

 in Sec. \ref{Sec3}. Results and
comments are detailed in the same section while the Sec.
\ref{Sec4} summarizes them and draws some perspectives.

\section{Coherent dynamical structure factor of the nonlinear KG lattice}\label{Sec2}

We assume that some light particles form a regular network,
whether it is inside a molecule or a crystalline solid. Were the
particle dynamics independent, the Hamiltonian would have read:
\begin{equation}
H_0= \sum_l\textsc{\Large [} \frac{p_l^2}{2m} + V(x_l)
\textsc{\Large ]} \label{Hsingle},
\end{equation}
where $x_l$ and $p_l$ are displacement and conjugate momentum
(i.e., $[x_l,p_l]=\hbar i$) of a particle at site $l$. For sake of
simplicity, we choose to work on a one-dimensional lattice with a
single direction for the motion of atoms, denoted by the unit
vector ${\bf u}$, which is named as polarization in the
followings. The single dimension proved relevant in several
concrete studies\cite{Fillaux2,Fillaux1,Kolesnikov,KolesZrH}. The
local potential $V(x_l)$ may then be expanded as a fourth order
series: $V(x_l)=a_2 x_l^2+a_3 x_l^3 +a_4 x_l^4$. Some higher order
terms could be added with no difficulties for our numerics, so as
to simulate eventually a specific shape of $V$. In truth, the
interaction between nearest neighbors, that may be direct or
mediated by the heavy atoms that compose the skeleton of a
material, involves a displacement coupling that is modeled by a
quadratic term. So the Hamiltonian of the particle ensemble now
reads:
\begin{equation}
H=H_0  - c \sum_{l,j=<l>} (x_l-x_{j})^2 \label{DebyeModif},
\end{equation}
where $c$ is the coupling parameter and the symbol $(<l>)$
describes the first neighbors of the site labeled by $l$.
Introducing the dimensionless operators $P_l$ and $X_l$, the
Hamiltonian can be reformulated as follows:
\begin{equation}
H = \hbar \Omega \sum_l \textsc{\Large[}\frac{P_l^2}{2} + A_2
X_l^2+A_3 X_l^3 +A_4 X_l^4 + \frac{C}{2} X_l \sum_{j=<l>} X_{j}
\textsc{\Large]} \label{Hamilton},
\end{equation}
where the fundamental frequency $\Omega$, as well as the
dimensionless coefficients $A_3$, $A_4$ and  $C$ have been defined
in Ref. \onlinecite{LPXarch}.
The total number of sites is denoted by $N$, the lattice parameter
by $a_0$ and the orientation of the chain is defined by a unit vector
${\bf v}$. We introduce $\psi$, the eigenstates of $H$, and
particularly the groundstate $\psi_{GS}$, the phonons
$\psi_{ph}({\bf k})$ and the biphonons $\psi_{bi}({\bf k})$. These
states are computed numerically\cite{LPXarch} as well as their
eigenvalues $E_{GS}$, $E_{ph}({\bf k})$ and $E_{bi}({\bf k})$,
respectively. The wave momentum, denoted by ${\bf k}$, verifies
${\bf k}= \frac{2\pi k}{N a_0} {\bf v} $ where $k$ is an integer.
The groundstate wave momentum ${\bf k}_0$ takes its value in the
reciprocal lattice, ${\bf k}_0= \frac{2\pi k_0}{a_0} {\bf v}$
where $k_0$ is an integer. Apart when it will be noted, $k_0$ is
fixed to zero. The lattice is assumed to be maintained at very low
temperature, i.e., $kT<E_{ph}$, so the
scattering induces some
excitations solely from $\psi_{GS}$. The transition
probability is proportional to the DSF\cite{Ashcroft}:
\begin{equation}
S({\bf q},\omega_{\psi({\bf k})})=\frac{1}{N}|\sum_j
<\psi_{GS}|e^{i [{\bf q}.{\bf r}_j]}|\psi({\bf k})>|^2\label{Phiq}
\end{equation}
where $\omega_{\psi({\bf k})}=(E_{\psi({\bf k})}
-E_{\psi_{GS}})/\hbar$, ${\bf q}$ is the scattering vector and
${\bf r}_j$ represents the atomic position at site $j$. That
position can be expressed as ${\bf r}_j= j a_0 {\bf v}  + {\cal L}
X_j {\bf u}$, where ${\cal L}=\sqrt{\hbar/(m\Omega)}$ is the
length scale of vibrations, fixed to a reasonable value of $3 \%$
of $a_0$. The polarization vector ${\bf u}$ is the principal axis
of atomic displacement around the equilibrium position $(j a_0
{\bf v})$. We now spell out our method for computing the DSF. As,
strictly speaking the Bloch identity does not hold for a nonlinear
lattice\cite{Ashcroft}, we propose a derivation which differs from
the conventional one. Since ${\cal L}$ is much smaller than $a_0$,
the exponential function in Eq. (\ref{Phiq}) can be expanded as a
Taylor series with respect to the weakest of its arguments, the
term proportional to $\cal L$:
\begin{equation}
S({\bf q}, \omega_{\psi({\bf k})})=\frac{1}{N}|\sum_j e^{i [{\bf
q.v}] j a_0}\times  \sum_p \frac{(i [{\bf q.u}] {\cal L})^p}{p!}
<\psi_{GS}|X_j^p|\psi({\bf k})>|^2.\label{Phili10}
\end{equation}
This development holds provided that the order of the expansion is
large enough. The series convergence
has been tested by increasing the order up to reach the desired
precision. The agreement to the standard analytical calculation on
the purely harmonic lattice (see Sec. \ref{Sec3}) is a necessary
condition of validity that has been fulfilled too. In the left
hand side of Eq. (\ref{Phili10}), the bracket
$<\psi_{GS}|X_j^{p}|\psi({\bf k})>$ can be replaced by $
<\psi_{GS}|X_0^{p}|\psi({\bf k})>$ multiplied by a phase factor
$e^{-i ([{\bf k.v}] a_0 j)}$, because of the translational
invariance. The sum over the subscript $j$ in Eq. (\ref{Phili10})
gives zero for all ${\bf q}$, aside from the wave vectors that
match the momentum conservation $[({\bf q}-{\bf k}).{\bf v}]=0$.
This condition, as well as the conservation of the energy are
supposed to be satisfied for a given state $\psi({\bf k})$. The
calculation of $S({\bf q},\omega_\psi)$ can then be reduced to the
determination of the bracket $D(\psi({\bf
k}),p)=<\psi_{GS}|X_0^{p}|\psi({{\bf k}})>$,
since one may check that:
\begin{equation}
S({\bf q},\omega_{\psi({\bf k})})=N \delta_{[({\bf q}-{\bf
k}).{\bf v}]} |\sum_p \frac{(i [{\bf q}.{\bf u}]{\cal L})^p}{p!}
D(\psi({\bf k}),p)|^2.
\end{equation}
We study the variation of the DSF along
a single direction which the angles
with respect to ${\bf u}$ and ${\bf v}$ are $\alpha_u$ and
$\alpha_v$, respectively. Then $S({\bf q},\omega_{\psi})$ depends
only on the magnitude of the momentum transfer $|{\bf q}|$ along
that direction and the conservation law fixes $|{\bf q}|
cos(\alpha_v) = [{\bf k}.{\bf v}]$. This condition can be achieved
whether $|\alpha_v|\neq \pi/2$ which may be reasonably assumed
for a low-angle scattering. In a same manner, when
$|\alpha_u|= \pi/2$ the inelastic part of the
structure factor is identically null
because $[{\bf q}.{\bf u}]=0$ (excepted for
the elastic DSF which then equals N). Consequently, we
assume that
$|\alpha_v|=|\alpha_u|\neq \pi/2$ which keeps the physics of our problem,
avoiding the unimportant cases. We
denote by $q$ the scalar product $ [{\bf q}.{\bf u}]$ for a vector
${\bf q}$ that matches
the conservation law, so that we have also
$q=[{\bf k}.{\bf v}]$. Then, the DSF reads:
\begin{equation}
S(q,\omega_{\psi(q)}) =N |\sum_p \frac{(i q{\cal L})^p}{p!}
D(\psi(q),p)|^2.\label{series}
\end{equation}

Since the optical modes are characterized by a weak value of $C$
in Eq. (\ref{Hamilton}), the first step of our treatment is
concerned with a perturbation theory with respect to the intersite
coupling. At $C=0$, namely the anti-continuum (e.g. , uncoupled,
atomic or molecular) limit\cite{Aubry}, the phonon states and the
phonon bound states can all be written as some Bloch
waves\cite{LPXarch}:
\begin{eqnarray} B_{\alpha} ({\bf k}) =
\frac{1}{\sqrt{N}} \sum_j e^{-i  [{\bf k.v}] a_0 j} \times
\phi_{\alpha,j} \Pi_{l\neq j} \phi_{0,l} \label{OSPBW}
\end{eqnarray}
where $\phi_{\alpha,j}$ is a on-site wave function that depends
only on $X_j$. Actually, $\phi_{\alpha,j}$ is the $\alpha$th
eigenstate of the on-site Hamiltonian:
\begin{equation}
h_l =  \frac{P_l^2}{2} + A_2 X_l^2+A_3 X_l^3 +A_4 X_l^4 .
\label{HamiltonOS}
\end{equation}
To a first order in $C$, the lattice groundstate $\psi_{GS}$ is simply
given by the product $B_{0}=\Pi_l \phi_{0,l}$ while Eq. (\ref{OSPBW})
gives a phonon for $\alpha=1$
and a biphonon for $\alpha=2$.
Let us denote by $V_{\alpha,n}$, the projection of the state
$\phi_{\alpha}$ onto the $n$th on-site harmonic oscillator
eigen-state $|n>$ and develop
the bracket $G(\alpha,0,p)=<\phi_{0}|X^p|\phi_{\alpha}>$:
\begin{eqnarray}
G(\alpha,0,p)=\sum_{n\geq 0,m\geq 0} V_{0,m} V_{\alpha,n} <{m}|X^p|n>\label{Galpha}
\end{eqnarray}
where the subscript has been dropped for ease.
We point out that for $C=0$, we have $D(B_\alpha({{\bf k}}),p)=
{G(\alpha,0,p)}/{\sqrt{N}}$ and $D(B_0,p)= G(0,0,p)$. One more
thing has to be done before reaching our goal, is to compute
$<{m}|X^p|n>$. To that purpose, the Bose-Einstein operators
[i.e. $a^{+}=\sqrt{2} (X-iP)$ and $a=\sqrt{2} (X +iP)$]
are used to expand $X^{p}$. We wrote a fortran program
which realizes the expansion, respecting the
commutation rule, $[a,a^+]=1$. For example, the output for
$p=10$ is:
\begin{eqnarray}
X^{10}&=&\frac{1}{32}[a^{+10}+a^{10}+10(a^{+}a^{9}+a^{+9}a^{})\nonumber\\
&+& 45(a^{+2}a^{8}+a^{+8}a^{2}+a^{+8}+a^{8})\nonumber\\
&+&120(a^{+3}a^{7}+a^{+7}a^{3})+360(a^{+}a^{7}+a^{+7}a^{})\nonumber\\
&+&210(a^{+4}a^{6}+a^{+6}a^{4})+1260(a^{+2}a^{6}+a^{+6}a^{2})\nonumber\\
&+&630(a^{+6}+a^{6})+
252a^{+5}a^{5}+2520(a^{+5}a^{3}+a^{+3}a^{5})\nonumber\\
&+&3780(a^{+5}a^{}+a^{+}a^{5})+3150(a^{+4}+a^{4}
+a^{+4}a^{4})\nonumber\\
&+&4725 (2a^{+2}a^{4}+2a^{+4}a^{2}+4a^{+2}a^{2}+a^{+2}+a^{2}
+2a^{+}a^{})\nonumber\\
&+& 12600(a^{+}a^{3}+a^{+3}a^{}+ a^{+3}a^{3})+945].
\end{eqnarray}
The writing of
$X^{p}$ would take more than one full page for
values of $p$ larger than $30$. Our program allows us to compute
the bracket $<{m}|X^p|n>$ up to the power $p=60$, for
different integers $m$ and $n$. The association of this program
with the numerical diagonalization of $h_l$ (Eq.
(\ref{HamiltonOS})) which fixes the coefficients $V_{\alpha,n}$ in
Eq. (\ref{Galpha})\cite{LPXarch}, permits to compute the
coefficients $G(\alpha,0,p)$ that can be tabulated for
different model parameters. Finally, to a first order in $C$, one
obtains:
\begin{equation}
S(q,\omega_{B_\alpha({q})})= |\sum_p \frac{(i  q {\cal L})^p}{p !}
G(\alpha,0,p)|^2  \label{jhk2}
\end{equation}
for the inelastic scattering
where the momentum conservation imposes $q={2\pi k}/{(N a_0)}$,
whereas the elastic response is given by:
\begin{equation}
S(q,0)= N |\sum_p \frac{(i  q {\cal L})^p}{p !}
G(0,0,p)|^2  \label{jhk22}
\end{equation}
where $q={2\pi k_0}/{a_0}$  [$k$ and $k_0$ range over integers].
Formally, these results are given for a one-dimensional lattice
but they can be extended to higher dimensional lattice by summing
over the coordinates and polarizations. To a first order in $C$,
the frequency $\omega_{B_\alpha(q)}$ can also be evaluated by:
\begin{equation}
\omega_{B_\alpha(q)}= \Omega (\gamma_\alpha-\gamma_0 -2C\times
G(\alpha,0,1)^2 \cos(q a_0))\label{EnFirstOrder}
\end{equation}
where the coefficient $\gamma_\alpha$ is the eigen-energy of $h_l$
(Eq. (\ref{HamiltonOS})) associated to $\Phi_\alpha$. For the
perfect harmonic lattice,
the first order in $C$ in the
standard calculation\cite{Lovesey,Ashcroft} of $S(q,\omega)$
is equivalent to
Eqs. (\ref{jhk2}) and (\ref{jhk22}).

When $C$ is large, the Hamiltonian in Eq. (\ref{Hamilton}) must be
diagonalized numerically after expanding in a suitable basis. We
worked with a Bloch wave basis\cite{LPXarch} given by:
\begin{eqnarray}
B_{[ \Pi_j \alpha_j]} ({\bf k}) = \frac{1}{\sqrt{A_{ [ \Pi_j
\alpha_j]}}} \sum_j e^{-i [{\bf k}.{\bf v}]a_0 j} \times \Pi_l
\phi_{\alpha_l,l+j}\label{OSPBW2}
\end{eqnarray}
where $A_{ [ \Pi_j \alpha_j]}$ ensures the normalization.
Each eigenstate, characterized by a wave vector ${\bf
k }$, can be written as a linear combination of those Bloch waves:
\begin{eqnarray}
\psi({\bf k })= \sum_{\Pi_j \alpha_j}  W_{\psi,\Pi_j \alpha_j}
B_{[ \Pi_j \alpha_j]} ({\bf k }),
\end{eqnarray}
where the subscript $\Pi_j \alpha_j$ identifies a single Bloch
wave in Eq. (\ref{OSPBW2}). The numerical diagonalization has been
carried out for different lattice sizes with no noticeable
discrepancy in the eigenspectrum in increasing $N$. In Fig.
\ref{fig0}, the same energy cutoff on the Bloch wave basis has
been fixed for both $N=23$ and $N=33$. Apart on their wave vector,
the eigenvalues are found to be independent of $N$. In Fig.
\ref{fig0} (a), the optical phonon branch is plotted whereas, in
Fig. \ref{fig0} (b), the energy region of the first overtone is
plotted for same parameters as in (a). One notes clearly the
biphonon branch and the two-phonon band, thoroughly described
elsewhere\cite{AGRANO,ScottLastChapter,bishop96,Eilbeck,LPXarch}.
Interestingly, for small enough nonlinear parameters, the
anharmonicity of the lattice is negligible  compared to the phonon
branch width and the biphonon branch disappears (see Fig.
\ref{fig0} (c)). The bracket $D(\psi,p)$ can now be written as
follows:
\begin{eqnarray}
D(\psi({\bf k }),p)= \sum_{(\Pi_j \alpha_j, \Pi_j \beta_j)}
W_{\psi_{GS},\Pi_j \beta_j}^* W_{\psi,\Pi_j \alpha_j} <B_{ [ \Pi_j
\beta_j]}(0)|X^p_0|B_{ [ \Pi_j \alpha_j]}({\bf k })> \label{Exact}
\end{eqnarray}
where the last term in the right hand side can be detailed
further:
\begin{eqnarray}
<B_{ [ \Pi_j \beta_j]} (0)|X^p_0|B_{ [ \Pi_j \alpha_j]}({\bf k
})>&=&\frac{1} {\sqrt{A_{ [ \Pi_j \alpha_j]}A_{ [ \Pi_j
\beta_j]}}} \sum_{n_1,n_2} e^{-i [{\bf k}.{\bf v}].a_0 n_1 }\nonumber\\
& & \times \Pi_{l\neq j }\delta(\alpha_{l+n_1},\beta_{l+n_2})
G(\alpha_{l+n_2},\alpha_{l+n_1},p).\label{Exact2}
\end{eqnarray}
This equation concludes our computation task on the KG lattice
dynamical response to the scattering of a external beam
of light or particles. Our
numerical treatment, including the diagonalization of $H$ and the
computations of Eqs. \ref{series}, \ref{Exact} and \ref{Exact2}
takes few hours on a conventional desktop PC. The convergence of
the series in Eq. (\ref{series}) has been tested by comparing our
results for different development orders, e.g. $30$, $40$, $50$
and $60$. As expected, the smaller is the scattering vector, the
better is the precision. For a momentum transfer that ranges over
$30$ lattice Brillouin zones, the difference between the DSF computed
with series orders $50$ and $60$ is less than 1 $\%$.




\section{Results and discussions}\label{Sec3}

The Fig. \ref{fig00} shows our typical result that is a 3D plot of
the inelastic $S(q,\omega)$ for a nonlinear lattice which model
parameters are those of Figs. \ref{fig0} (a) and (b). Three
noteworthy resonances emerge with same order of magnitude. These
three resonances correspond to the eigen-energies of the nonlinear
phonon states, i.e., the phonons, biphonons and triphonons. The
binding energy of the biphonon, evaluated at the center of the
Brillouin zone (BZ) is around a few percent of the fundamental
phonon excitation, as measured in different metal
hydrides\cite{Kolesnikov,KolesZrH}. Several model parameters have
been tested and lead to qualitatively similar results. For
instance, besides the different values of the energy transfer, the
biphonon and triphonon resonances occur in a lattice with a
quadratic-quartic on-site potential, i.e., $A_3=0$ in Eq.
\ref{Hamilton}. In the biphonon and triphonon resonances, one may
recognize some peaks that will be examined further. In order to
analyze our results more quantitatively it is, though, easier to
work with a $2$D plot that represents the $S(q,\omega)$ profile
versus the projection $q=[{\bf q}.{\bf u}]$ of the scattering
vector ${\bf q}$. In order to verify the accuracy of our
computations, presented in Sec. \ref{Sec2}, we compare our
numerical results to the exact analytical ones that are achieved
in a purely harmonic lattice. In that case, the profile of the
Debye-Waller factor, i.e., $S(q,0)/N$ is plotted in Fig.
\ref{fig1} and shows a convincing agreement since our numerical
results scarcely differ from the analytical ones. It ensures us
that our series expansion of $S(q,\omega)$ has converged. Only the
non-zero values of the Debye-Waller have been retained in the
plot, i.e., $q={2\pi k_0}/{a_0}$. A similar accord is obtained for
the inelastic part of $S(q,\omega)$ (see further in the same
section). The calculation made in a standard harmonic
approximation\cite{Ashcroft} gives a Gaussian dependency of the
Debye-Waller factor $S(q,0)/N=\exp(-2 W(q))$  where $2W(q)=(q{\cal
L})^2 |<\psi_{GS}|X^2_j|\psi_{GS}>|$. In this expression, the
bracket is simply given by $1/(2N)\sum_K 1/\omega_{ph}(K)$ where
$\omega_{ph}(K)=\sqrt{1+2C cos(K.a_0)}$ and $K$ ranges in the
first BZ. When the nonlinear parameters are no longer negligible
in the Eq. (\ref{Hamilton}), we note, in Fig. \ref{fig1} that the
Debye-Waller of a nonlinear lattice differs from the harmonic one.
The Gaussian form is yet roughly conserved so there is no
qualitative changes involved by the on-site anharmonicity. Our
perturbation theory proves sufficient under the condition that the
inter-molecular coupling is not too large. In agreement with our
results, the earlier study of B.V. Thompson\cite{Thompson}
concluded that a cubic-anharmonic term involved a negligible
correction upon the Debye-Waller factor. The calculation of
Thompson is, in fact, a second order perturbation in the cubic
term, which is thus assumed to be weak enough for the perturbation
theory to be valid. Not surprizingly, it leads to a correction
that is also weak. However, as we shall see, the most significant
contribution of the nonlinearity to the DSF spectra proves to be
inelastic and is due to the phonon bound states.

In a same manner as for the elastic scattering, our numerical
computation of the inelastic DSF  is compared to the exact
calculation\cite{Ashcroft} in the case of a harmonic lattice. The
contribution of phonon states is then given by:
\begin{equation}
S(q,\omega_{ph}(q))=\frac{q^2}{2\omega_{ph}(q)}\exp{(-2W(q))},
\end{equation}
with same notations as previously used in the same section. The
obtained agreement, demonstrated in Fig. \ref{fig1bis}, confirms
the validity of our theoretical approach and consequently the
convergence of our series development. The DSF profile appears as
being continuous because the lattice size is large enough to blur
the discreteness of the Fourier space at the scale of Fig.
\ref{fig1bis}, which ranges over 60 BZ. Apart from the
$S(q,\omega)$ ripple, the perturbation theory captures quite well
the main variations of the envelop of $S(q,\omega_{ph})$. This
ripple stems from the dispersion of $\omega_{ph}$ and so, its
amplitude increases with $C$ to become some peaks for the acoustic
phonons in a harmonic lattice\cite{comment}. When the sign of $C$
is inverted in our model ($C$ becomes negative), the local minima
of $S(q,\omega_{ph})$ are shifted at the edges of the lattice
Brillouin zones instead of being in the middle when $C>0$. In Fig.
\ref{fig1bis}, the DSF resonance involved by the two-phonon states
is also plotted. It is usually termed multiple-scattering and is
one order of magnitude smaller than the one-phonon process. If the
anharmonicity of $V$ can not be neglected, the profile of the
one-phonon resonance in $S(q,\omega)$ shows no qualitative changes
compared to the harmonic lattice (see Figs. \ref{fig1bis} and
\ref{fig3} (a)). However, the response of the biphonon, which
profile is denoted by $S(q,\omega_{bi})$, appears clearly at
higher energy transfer than one-phonon (see Figs. \ref{fig00} and
\ref{fig3} (a)). The resonance associated to the biphonon is one
order of magnitude larger than for the unbound phonon states and
has same order as $S(q,\omega_{ph})$. A gap is opened between the
biphonon branch and the two-phonon band (Fig. \ref{fig0} (b)) so
the biphonon resonance occurs for an energy transfer which does
not match the harmonics of the fundamental phonons. As mentioned
above, that energy gap is also called the binding energy of the
biphonon\cite{AGRANO}. It is also worth noting that the maximum of
the $S(q,\omega_{bi})$ envelop is reached for a scattering vector
that differs from the envelop maximum of $S(q,\omega_{ph})$. This
maximum occurs, indeed, at larger $q$ for the biphonon. The
$S(q,\omega_{bi})$ envelop can be evaluated within the
perturbation theory (see Figs. \ref{fig3} (a)) which provides a
rather satisfactory approximation, although the $S(q,\omega_{bi})$
ripple has a larger amplitude than for phonons. To interpret our
results, it is needed to dwell on the ripple of the biphonon
contribution to the DSF. As for the phonon, this ripple is related
to the dispersion of the biphonon branch. In the case where the
binding energy of biphonon is not very large, say no more than a
few percent of the phonon energy, the different bottoms of the
$S(q,\omega_{bi})$ ripple occur at center of the successive BZ
while the tops are reached at the BZ edges. According to several
tests, it is a systematic feature which, in contrast to phonon,
does not depend on the sign of $C$. Moreover, as shown in the
insets of Figs. \ref{fig3} (a) and (b), the minima of
$S(q,\omega_{bi})$ and the maxima of the two-phonon resonance
correspond one to one. That may be explained as follows. In our
perturbation theory the biphonon states are approximated by the
Bloch waves that bear a single on-site excitation, $\alpha=2$ in
Eq. (\ref{OSPBW}). In case $C\neq 0$, since the biphonon binding
energy is not very large, these Bloch waves are hybridized with
some other Bloch waves, given by Eq. (\ref{OSPBW2}), and
particularly those bearing two on-site excitations $\alpha=1$, at
different sites. The degeneracy-lifting of the latest states
yields the unbound two-phonon band. The ripple of the DSF can not
be analyzed through our first order perturbation theory (see Figs.
\ref{fig3} (a) and (b)), which hints that it is due to the Bloch
waves hybridization, involved by the intersite coupling. The
discrepancy of the perturbation theory Eq. (\ref{jhk2}) increases
as the biphonon gap decreases (compare Figs. \ref{fig0} (b) and
(c) to Figs. \ref{fig3} (a) and (b)) because of the hybridization
step-up. The gap between the biphonon branch and the unbound
two-phonon band is smaller at the center of each BZ (see Fig.
\ref{fig0} (b) and Ref. \onlinecite{LPXarch}) because here the
width of the two-phonon band is maximum. The smaller the energy
gap is the larger the hybridization is, so as a result, one finds
a larger contribution of the two on-site excitation Bloch waves
into the biphonon eigenstate at the center of each BZ. The Bloch
waves with multiple excitations at distinct sites yield a zero
response to scattering, $S(q,\omega)=0$. One deduces that the
biphonon response $S(q,\omega_{bi})$ is minimum when the
contribution of the two on-site excitation Bloch waves is maximum,
i.e., at the center of each BZ. On the other hand, the unbound two
phonons resonance comes to a maximum at the center of each BZ
because of the contribution of the Bloch waves with a single
on-site excitation $\alpha=2$ to the two-phonon eigenstates.
Conversely, at the edge of the lattice Brillouin zones, the Bloch
wave hybridization is minimum (because the energy gap is maximum)
so that the biphonon response is maximum and may even reach the
value computed within the perturbation theory (Fig. \ref{fig3}
(b)). Another interesting point that is made clear within the
previous discussion is why the $S(q,\omega_{bi})$ ripple is not
shifted when the sign of $C$ is changed, unlike
$S(q,\omega_{ph})$. Indeed, reversing the $C$ sign leaves
unchanged the two-phonon band shape, i.e., the band width is still
maximum at the BZ center.
So it is for the biphonon branch which the dispersion
is mainly due to the contribution of the two
on-site excitation Bloch waves\cite{LPXarch}
(the dispersion computed within Eq. (\ref{EnFirstOrder})
for $\alpha = 2$ is much smaller than the effective
biphonon band width).
So the gap
between the biphonon and two-phonon bands is equally unchanged
under the inversion of the coupling sign, i.e.,  the energy
gap is still maximum at the BZ edges and minimum at the center.
According to our analysis of the $S(q,\omega_{bi})$ ripple, it
makes clear why the maxima and minima of the
biphonon DSF are independent of the sign of $C$.
Our argumentation holds provided that the biphonon binding energy
remains inferior to a few percent of the fundamental excitations.

When the anharmonicity of $V$ is weak compared to the phonon band
width, the biphonon gap closes and a pseudogap forms\cite{LPlett}
at the edge of the lattice Brillouin zones. Then the biphonon
branch hardly appears nearby the two-phonon band (see Fig.
\ref{fig0} (c)). The lattice is said quasi-harmonic since the
spectrum anharmonicity vanishes. There is, indeed, no anharmonic
resonances, neither in the density of state nor, accordingly, in
the DSF. That may happen for higher order phonon bound states as
the hardening of the quartic on-site term in $V$ (Eq.
(\ref{Hamilton})) could compensate the softening of the cubic
potential in a certain range of energy. Although that seemingly
perfect harmonicity, the $S(q,\omega)$ ripple may yet betray the
nonlinearity of $V$. In truth, at the BZ center, the DSF
associated to the biphonon, referred to as $S(q,\omega_{bi})$, has
same order as the DSF of two-phonon states, which magnitude is one
order below the phonon resonance (see Fig. \ref{fig3}(b)). In
contrast, the profile $S(q,\omega_{bi})$ reaches its maxima where
the pseudogap opens (see Fig. \ref{fig0} (c)), i.e., at the BZ
boundaries and the ripple alongside each BZ can be seen as a
series of peaks. In Fig. \ref{fig3} (b), the largest peak remains
much larger than the two-phonon response and it has same order as
the maximum of the phonon DSF profile, $S(q,\omega_{ph})$. The
maxima of the biphonon DSF decrease as the nonlinear parameters
tends to zero. However, before to reach that point, we have seen
that the anharmonicity of the energy spectrum vanishes as in Fig.
\ref{fig0} (c). Consequently, we propose to dub the
$S(q,\omega_{bi})$ peaks by nonlinear Bragg peaks, since they
gather two features that are opposite to the usual Bragg
scattering: first, they are inelastic peaks and second, they
satisfy the Bragg reflection condition but shifted of a half
reciprocal lattice vector (i.e., $\pi/a_0$ for the one-dimensional
chain). According to Ref. \onlinecite{LPlett}, the pseudogap may
occur in lattices with a higher dimension so that similar results
as those presented here can be expected in different lattices. In
spite of the relative simplicity of our model compared to the case
encountered in practice, it indicates that the biphonon may still
emerge and contribute significantly to the scattering, even though
the spectrum seems perfectly harmonic. In Figs. \ref{fig0} (c) and
\ref{fig3} (b), we have chosen intentionally a set of parameters
that yields a eigen-spectrum which is almost harmonic. A pseudogap
hardly opens between the biphonon branch and the two-phonon band.
The parameter $A_3$ could be doubled [as in Fig. 2 (c) in
Ref.\onlinecite{LPlett}] without opening a substantial gap which
would permit to identify the anharmonicity in the energy density
of state. In fact, the larger is the phonon dispersion the larger
is the range of nonlinear parameters that leads to a
quasi-harmonicity. In Fig. \ref{fig4} (a) the profile of
$S(q,\omega)$ is plotted for different values of the energy
transfer $\omega$, e.g., $\omega_{ph}$ for phonon, $\omega_{bi}$
for biphonon and $\omega_{tri}$ for triphonon. The triphonon
resonance is found to reach a significant magnitude that is
comparable to phonon and biphonon. Such a high order phonon bound
state has been identified in different materials, e.g.,
TiH\cite{Kolesnikov} and ZrH\cite{KolesZrH} and HCl\cite{Gellini}.
It is worth noting the triphonon spectrum ripple, which originates
from the hybridization between the bound and unbound phonon
states, as for the biphonon spectrum. This triphonon ripple is
shifted of $\pi/a_0$ compared to the biphonon DSF profile.

In what follows, we spell out how our theory could be linked to
some concrete cases. To simulate a scattering experiment, we must
account for the resolution of the technics. If one assumes that
the energy resolution is not sufficient to differentiate the
inelastic lattice resonances (as it may be the case, for instance,
in the inelastic X-rays scattering), we have to integrate our
simulation with respect to the energy transfer $\omega$. The
elastic contribution is skipped from that integral so as to
distinguish the inelastic part of the DSF. We computed the
corresponding integral and reported the result as a dashed line in
Fig. \ref{fig4} (a). The amplitude and the position of the maximum
of the DSF integral are related to the nonlinear parameters. In a
strictly harmonic lattice, that integral scarcely differs from the
phonon response, whereas in Fig. \ref{fig4} (a) we note a clear
difference between the dashed line and the profile
$S(q,\omega_{ph})$. The envelop of the DSF integral has a form
that might be fitted by a superposition of two Gaussian functions,
centered at different momenta. It is a standard treatment in the
interpretation of the experimental $S(q,\omega)$ profile (see for
instance Fig. 8 in Ref.\onlinecite{Fillaux2}). If the nonlinear
Bragg peaks are large (i.e., when the anharmonicity is weak
compared to the phonon dispersion but that is yet not negligible)
they may emerge in the integral as shown for large $q$ in Fig.
\ref{fig4} (a). In a strongly nonlinear lattice, a large gap
separates the biphonon branch from the two-phonon energy band so
the nonlinear Bragg peaks shrinks to a ripple. Then it becomes
rather difficult to depict the contribution of the nonlinear
states in the DSF integral, aside from the shift of the envelop to
larger ${\bf q}$.
On the other hand, if the energy resolution is sufficient to
resolve the energy dispersion of the phonon branch but the accuracy over
the scattering vector is not to separate the BZ
(as it may be the case, for instance, in the neutron
scattering of a powder specimen, since ${\bf u}$ and
${\bf v}$ are randomly distributed), then our simulation must account
for the contribution of the distinct BZ.
As in a one-dimensional lattice the
only symmetry is the inversion, we have to sum
the factor $2S(q+k_0,\omega(q))$ over the reciprocal vector $k_0$,
where $q$ and $\omega(q)$ are both assumed to match the momentum and
the energy conservation, respectively.
Since $S(q+k_0,\omega(q))$ decreases exponentially
for a large enough value of $(q+k_0)$ (see Fig. \ref{fig4} (a)), the sum is
ensured to be finite. The result of our treatment is shown in
Fig. \ref{fig4} (b). We note the three resonances due to the
nonlinear states, i.e., the biphonon, the triphonon and the quadriphonon
that are separated from the broad resonances of either the phonon or
the unbound phonons.
Here, we must enhance that whether the
inelastic scattering is coherent or incoherent, the lattice
resonances occur at same energy transfer. Thus, from that point of
view it is relevant to compare our simulation to the experimental
spectrum of incoherent INS in metal hydrides. Actually, the model
parameters in Fig. \ref{fig4} (b) has been adjusted so as to
obtain the same energy resonances for the phonon, biphonon and
triphonon as in the spectrum shown in Fig. 2  of Ref.
\onlinecite{KolesZrD}. The energy unit is $\hbar \Omega\approx 107
meV$, the dimensionless coupling is $C=0.06$ and the nonlinear
parameters are $A_3=0.17$ and $A_4=0.0231$. In that manner, the
on-site nonlinearity and thus the energy landscape of the light
particles, here the deuterium, may be worked out to a better
accuracy than in a purely quadratic model (see the inset in Fig.
\ref{fig4} (b)). The same simulation could have been achieved in
other alloys as TiH and ZrH\cite{Kolesnikov,KolesZrH}. In their
earlier studies, A.I. Kolesnikov and
co-workers\cite{Kolesnikov,KolesZrH,KolesZrD} recognized the
phonon bound states resonances in the INS spectra of TiH-D and
ZrH-D. To analyse their experimental
spectra, the authors compared their data
to a simulation made with a model proposed initially by
V.M. Agranovich\cite{AGRANO}, i.e., a one-dimensional boson
Hubbard lattice. In such a model it is, though, difficult to
figure out the potential landscape of the interstitial gaps.
The single dimension of the model was shown to be reasonable
because of the strong anisotropy of the H-H interaction along the
metal c-axis. Since we are studying the coherent DSF, a accurate
simulation of the spectra obtained by Kolesnikov {\it et al.} for
the incoherent INS is out of purpose. We shall attempt such a
exercise in a future work, devoted to the computation of the
incoherent DSF. The amplitudes of various resonances as well as
their widths might then be inferred. In Fig. \ref{fig4} (b), our
simulation has been realized with a typical energy resolution
function, i.e., a triangle function centered at $\omega$ with a
width of $0.02 \omega$, which corresponds to the crystal analyser
TFXA, ISIS Rutherford Appleton Laboratory, described in Ref.
\onlinecite{KolesZrH}. The dynamical trajectory of the TFXA
machine has not been simulated in Fig. 6 (b). It is noteworthy
that the amplitude of the nonlinear resonances has same order as
for phonons. Similar calculations with different model parameters
even showed that, for a stronger anharmonicity the nonlinear
resonances may be even larger in magnitude. The reason for such a
behavior is that the density of state enhances as the eigenstates
band width decreases. This effect is sharp around the narrow
branches as those of phonon bound states\cite{note2} (see Fig.
\ref{fig0} (b)) which explains why the biphonon resonance may
dominate the phonon one. The increase in the density of state may
be recognized too in the single phonon response because the phonon
branch is flat at the edge and at the center of the lattice BZ
(see Fig. \ref{fig0} (a)). Thus one sees two side-bands in the
phonon resonance, at the upper and lower boundary in energy. These
side-bands may be identified as phonon wings. According to our
reading of the above cited works, the experimental INS spectra of
the metal hydrides do not exhibit that property but, once again,
one must left that point aside, until we study the incoherent DSF.
On the other hand, it is worth noting that the form of the lattice
response around the first and the second overtone energy regions,
in Fig. \ref{fig4} (b), is similar either to the infrared
adsorption spectra\cite{bogani2} in crystals as CO$_2$, N$_2$O and
OCS  or to the Raman spectrum of H$_2$ solid\cite{Mao}. Indeed the
emergence of a sharp peak, associated to a biphonon (or to a
bivibron\cite{Mao}) occurs near a small hill-like resonance that
is due to the unbound phonon states. In our simulation, a gradual
variation of the model parameters so as to decrease the
anharmonicity shows the same behavior as the pressure-induced
bound-unbound transition, at $25$ GPa in H$_2$ solid\cite{Mao}, or
at $34$ GPa in D$_2$ solid\cite{MaoD2}. Around that transition,
the biphonon (or bivibron) peak broadens and decreases in
magnitude. In our model, the pressure variation can be simulated
by a change of the coupling parameter $C$ due to the fact that
neighbouring molecules are moved closer together because of the
external pressure.


Although our work is concerned with a one-dimensional lattice,
the results enhanced in the present study
should hold in higher dimension where
the DSF depends on the lattice symmetries and polarizations.
The extension to higher dimension could be achieved
in our perturbation theory
with no particular difficulty. This
approach proves adequate under the condition that
the intersite coupling is very weak,
as expected in the metal hydrides as
NbH and TaH, from the study of
J. Eckert {\it et al.}\cite{Eckert}. Our numerics, more accurate,
could also be extended to higher dimension but that would require
either to use a powerful computer or else a dramatic
restriction on the site number, which could yet be relevant for
some molecules as benzen\cite{benzen}.

\section{Conclusion}\label{Sec4}
The dynamical structure factor (DSF) of the nonlinear Klein Gordon
chain (KG) has been calculated for different strength of the
on-site anharmonicity. This has been possible thanks to our
numerics that permit to diagonalized accurately the non-quadratic
Hamiltonian\cite{LPXarch}. The DSF has been expanded as a Taylor
series of the atomic displacements, which avoids the use of the
conventional harmonic approximations. We found that the on-site
nonlinearity leads to phonon bound states and consequently to some
anharmonic resonances in the DSF spectrum, which somehow confirms
other works on different quantum
lattices\cite{bishop96,AGRA1976,Dubovski1997}. The treatment of
the intersite coupling in a perturbation theory proved
satisfactory to tackle a lattice which nonlinearity is stronger
than the dispersion. In contrast, when the interaction between
first neighbors dominates the on-site nonlinearity, which remains
however non-zero, the amplitude of the dynamical structure factor
makes visible the biphonon, although the energy transfer is almost
harmonic. In such a case, the variation of the biphonon DSF with
respect to the transfer momentum $q$ exhibits some peaks that are
much larger than the multiple-scattering. This would hint that,
provided that the DSF could be resolved accurately in $q$, the
nonlinear behavior of certain materials could be worked out, even
though their spectra show no apparent energy anharmonicity.

In certain metal hydrides whose hydrogen anharmonicity is
significant, we proposed a scheme to work out from the INS
spectra, the potential landscape of the interstitial gaps. To
approach some concrete cases, the theory requires, thought,
further developments to consider, for instance, the three
dimensionality of a realistic sample or the incoherent scattering.
In a first step, this could be achieved within our perturbation
theory. In addition to the possibility of simulating the
incoherent INS in certain hydrides and molecular crystals, as
those quoted in the present paper, it might be worth studying the
nonlinear surface modes that could be investigated practically by
different technics, e.g., electron scattering or infrared
adsorption. The low dimensionality of the surface could give an
advantage to achieve a simulation that would be physically
accurate, particularly upon the lattice geometry and various
polarizations. As an example, we note the case of the CO molecules
adsorbed on a Ru($111$) surface that has been studied both
theoretically and experimentally\cite{Jakob,Bonn,Pouthier}.
As a low dimensional system, a vicinal surface, which might
exhibit a regular one-dimensional nano-structure over several
micrometers, would be the ideal substrate to explore the coherent
scattering of a appropriate nonlinear surface modes. The behavior
of molecules adsorbed on a vicinal surface presents some
specificity (see for instance Refs.\onlinecite{Marinica,Girardet}
and references therein) and thus the feasibility of such a study
remains under question. Though, the aim would be to carry out a
momentum resolved experiment, as those attempted on the PtCl
ethylene diamine chlorate\cite{mackay2000,ISIS}, to check whether
the predictions established within the KG model are confirmed in
practice, especially about the biphonon contribution to the DSF.
Another possibility to test our theory, would be on the O$_2$
solid\cite{Horl,mackay2000} where the oxygen cross-section is
mainly coherent. Although, in the $\beta$-phase of the O$_2$
crystal, the inter-molecular coupling overpasses the anharmonicity
of the O$_2$ stretch\cite{Horl}, our study showed that in such a
case the inelastic scattering due to biphonon could yet be larger
in magnitude than the two-phonon response. Further, the phonon
bound states could emerge at higher energy transfer as found for
the triphonon and quadriphonon in metal hydrides\cite{KolesZrD}.
At least from a theory viewpoint, the $\beta$-phase of the O$_2$
solid might be worth reexamining with recent experimental devices.
Finally, according to the authors of Ref.\onlinecite{Ross1998},
the potential landscape of hydrogen in PdH, that has been worked
out from the INS spectra\cite{Ross1998} agrees well with a
first-principles calculation made independently\cite{Elsasser}. On
the basis of that successful comparison, one may expect that a
standard first-principles theory could be used to calibrate the KG
parameters so as to simulate ab-initio the dynamical structure
factor.

\begin{acknowledgments}
I gratefully acknowledge Robert S. MacKay who invited me to attend
a neutron scattering experiment\cite{ISIS} at Rutherford Appleton
Laboratory, which motivated the start of the present study.
I also thanks M. Guttmann who pointed the
metal hydrides as a possible concrete example for the theory.
\end{acknowledgments}

\newpage

\centerline{Figure captions}

\begin{figure}
\noindent
\includegraphics[width= 9cm]{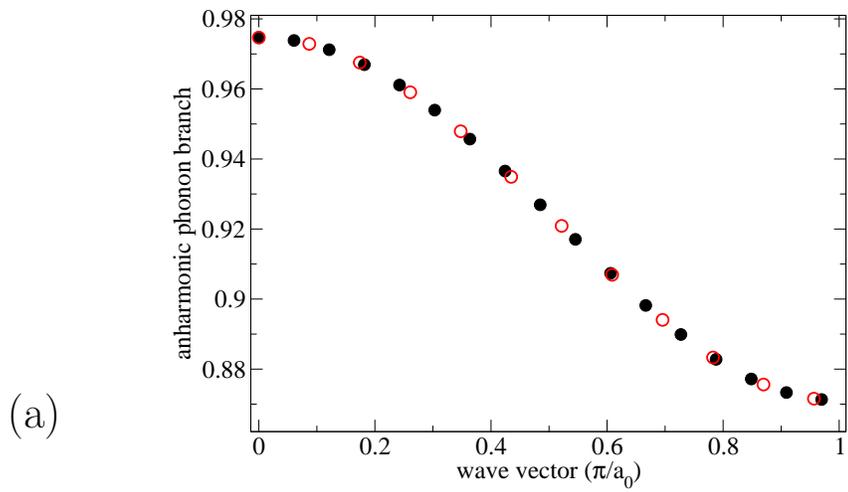}\\
\vspace{1.5cm}
\includegraphics[width= 9cm]{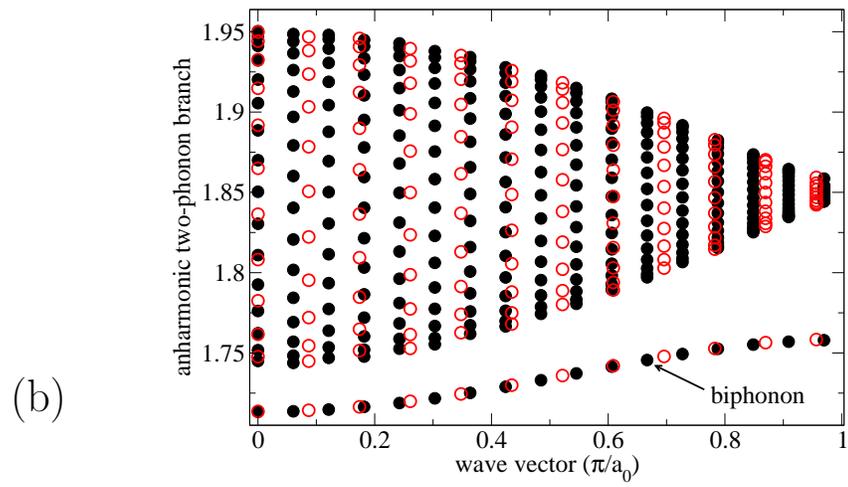}\\
\vspace{1.5cm}
\includegraphics[width= 9cm]{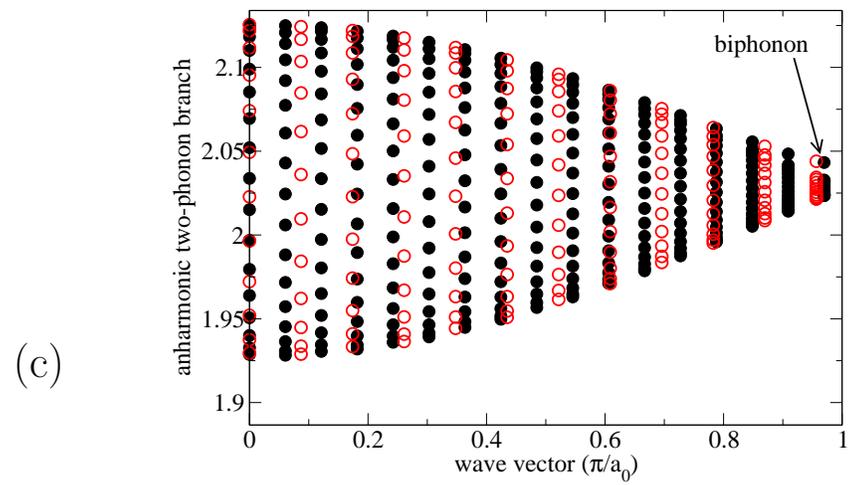}
\caption{\label{fig0} \bf{(2005) L. Proville}}
\begin{picture}(300,10)(0,0)
\put(-30,530){\makebox(0,0){\Large (a)}} \put(-30,310){\makebox(0,0){
\Large (b)}}\put(-30,100){\makebox(0,0){
\Large (c)}}
\end{picture}
\end{figure}

{Fig. \ref{fig0}:
Energy spectrum of a chain composed of $N$ unit cells,
$N=33$ (full circles)
and $N=23$ (empty circles):
(a) the optical phonon branch for model parameters $C=0.05$,
$A_3=0.12$ and $A_4=0.01$; (b) the two-phonon energy region for
the same parameters as (a); (c) the same as in (b) but for
$C=0.05$, $A_3=0.05$ and $A_4=0.01$.
The Y axis unit is $\hbar \Omega$.
The wave vector is reported on the X axis,
whose unit is $(\pi/a_0)$ and ranges over
the lattice first Brillouin zone.}\\

\begin{figure}
\noindent
\includegraphics[width= 18cm]{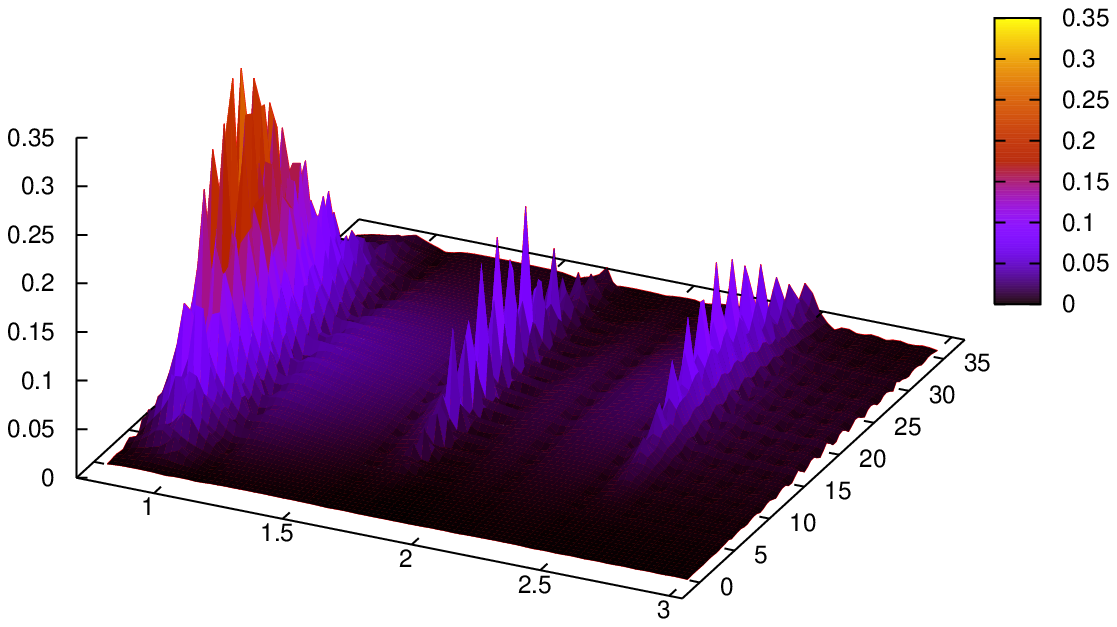}
\caption{\label{fig00} \bf{(2005) L. Proville}}
\begin{picture}(300,10)(0,0)
\put(65,90){{\Large $\omega$}} \put(305,120){{
\Large q}}\put(-50,180){\begin{rotate}{90}
{\large inelastic $S(q,\omega)$}
\end{rotate}}
\end{picture}\end{figure}

{Fig. \ref{fig00}:  (Color online) A 3D plot of the inelastic
$S(q,\omega)$ as a function of the dimensionless energy transfer
$\omega$ and the scalar product $q=[{\bf q.u}]$ of the transfer
momentum ${\bf q}$ and the polarization ${\bf u}$. The parameters
are the same as in Figs. \ref{fig0} (a) and (b). The figure is easier to depict in color.}\\

\begin{figure}
\noindent
\includegraphics[width= 13cm]{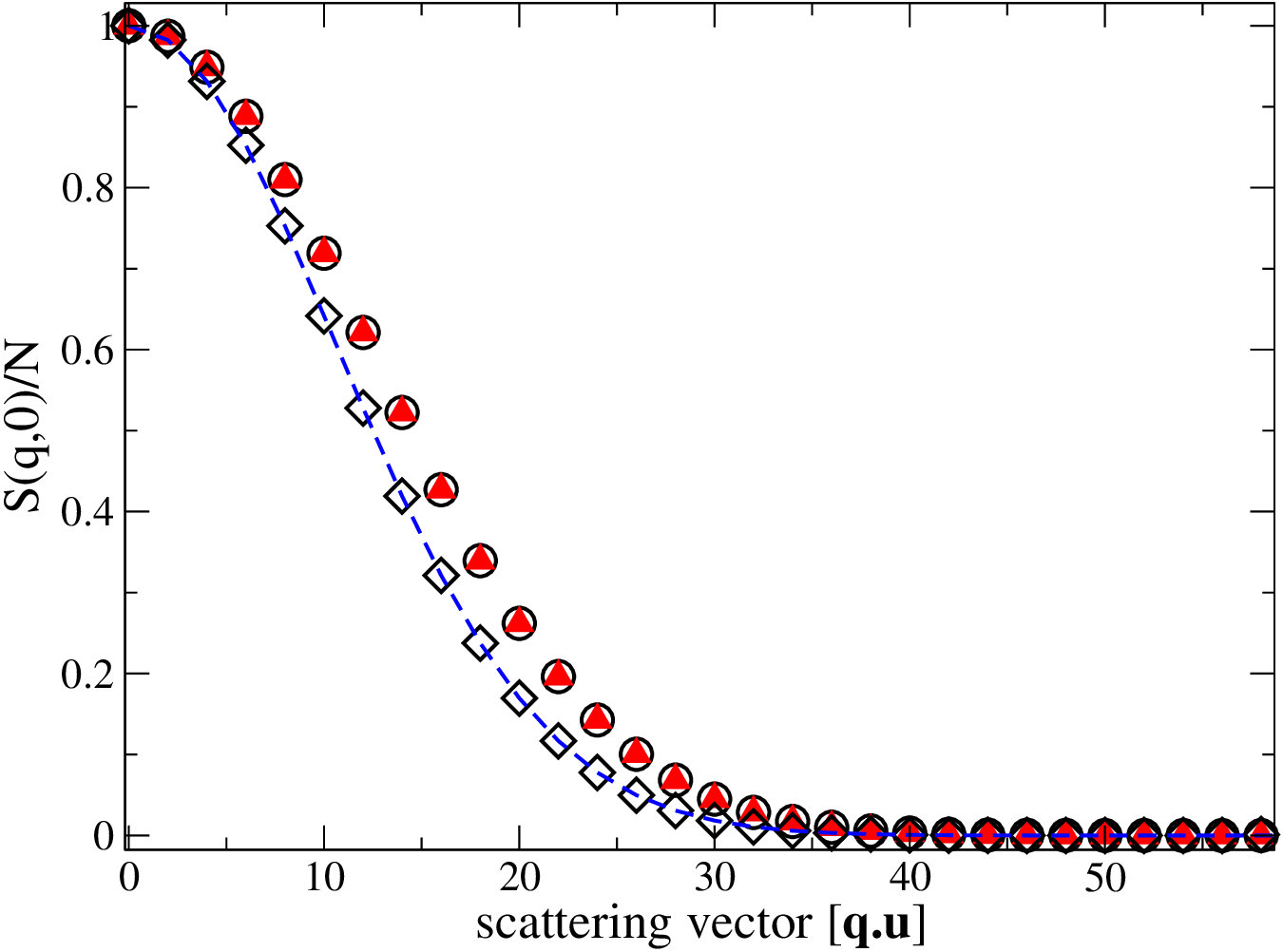}\\
\vspace{2cm}
\caption{\label{fig1} \bf{(2005) L. Proville}}
\end{figure}

{Fig. \ref{fig1}:  A plot of the
Debye-Waller factor $[S(q,\omega=0)/N]$ versus the scalar product
$q=[{\bf q.u}]$, for
different model parameters.
Each symbol corresponds to a Bragg peak.
Our numerical results are reported
for $C=0.05$, $A_3=A_4=0$ (diamonds) and
$C=0.05$, $A_3=0$, $A_4=0.2$ (triangles).
The exact formula derived in a
harmonic model (dashed line) confirms the former while
the latter has been  computed within a perturbation
theory (circles) too.
The X axis, which unit is $\pi/a_0$
bears the scattering vector $q$
over a range of $60$ Brillouin zones.}\\

\begin{figure}
\noindent
\includegraphics[width= 13cm]{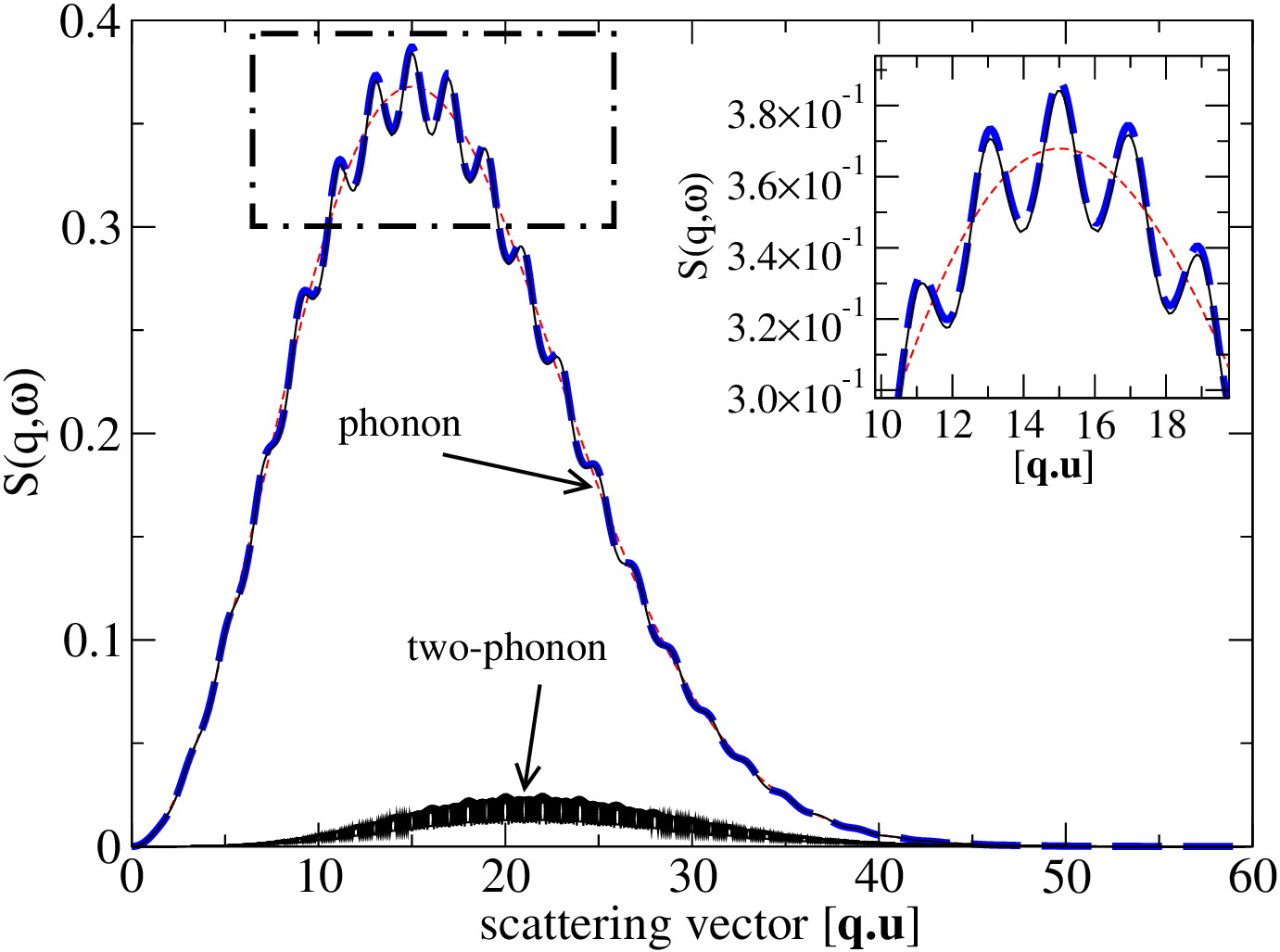}\\
\caption{\label{fig1bis} \bf{(2005) L. Proville}}
\end{figure}

{Fig. \ref{fig1bis}:  A plot of the coherent
inelastic $S(q,\omega)$ profile versus $q=[{\bf q.u}]$, for phonon in
a harmonic lattice: $C=0.05$, $A_3=A_4=0$. The thin
dashed line has been obtained by
a perturbation theory. The thin solid
line corresponds to our numerical treatment and the thick dashed
line to a plot of the exact formula\cite{Ashcroft} (the two latest curves
are hardly distinguished in the graph). In the inset, the area
in the dot-dashed rectangle
is magnified. The X axis unit is $\pi/a_0$.}\\

\begin{figure}
\noindent
\includegraphics[width= 12cm]{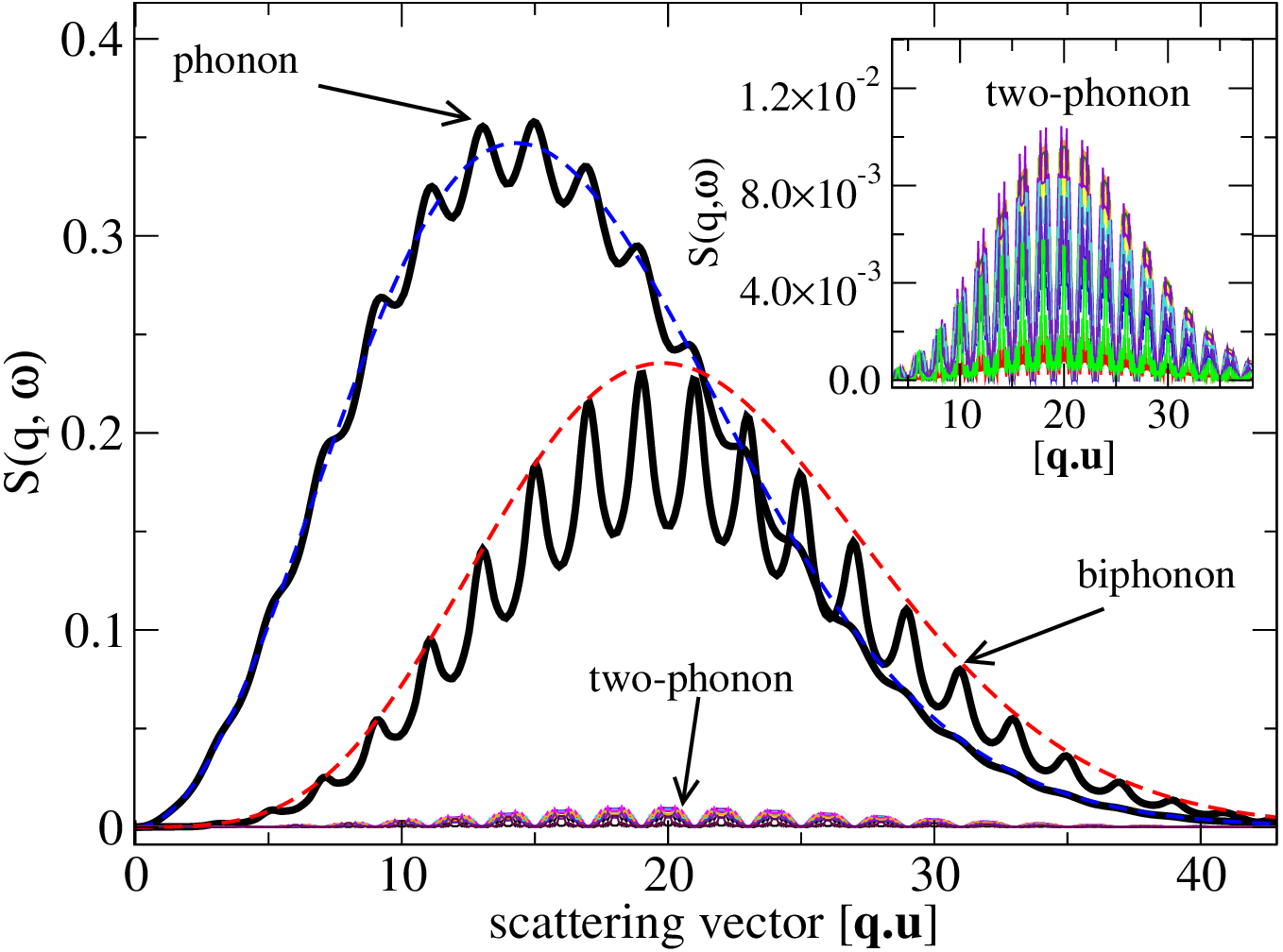}\\
\vspace{2cm}
\includegraphics[width= 12cm]{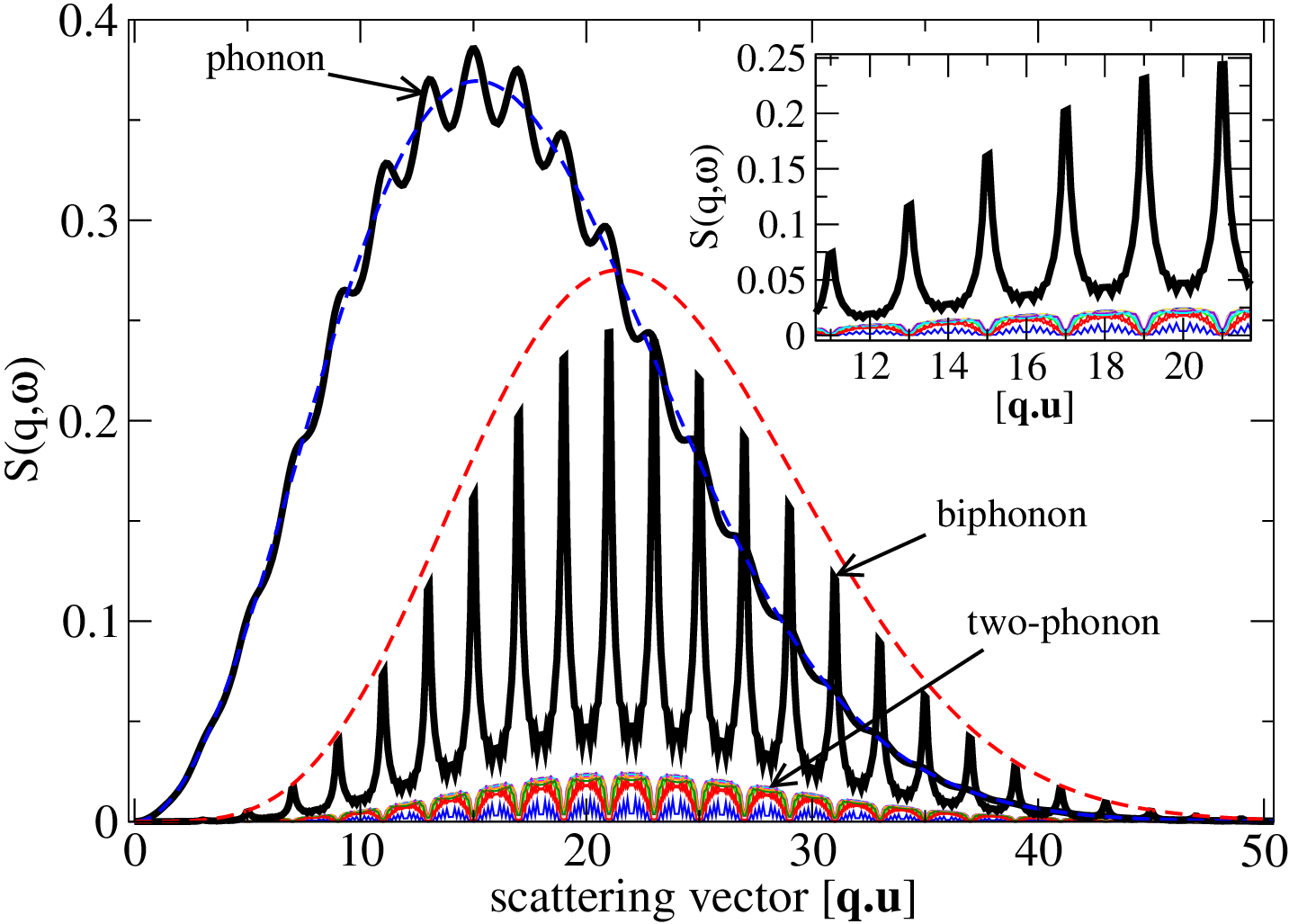}
\caption{\label{fig3} \bf{(2005) L. Proville}}
\begin{picture}(300,10)(0,0)
\put(-40,400){\makebox(0,0){\Large (a)}}
\put(-40,80){\makebox(0,0){\Large (b)}}
\end{picture}\end{figure}

{Fig. \ref{fig3}: A plot of the coherent inelastic $S(q,\omega)$
profile versus $q=[{\bf q.u}]$, for phonon, biphonon and
two-phonon states. The model parameters are: (a) $C=0.05$,
$A_3=0.12$ and $A_4=0.01$; (b) $C=0.05$, $A_3=0.05$ and
$A_4=0.01$. The dashed lines correspond to our perturbation theory
and the solid lines to our numerical treatment. The X axis unit is
$\pi/a_0$ and the scattering vector $q$ ranges over $60$ Brillouin
zones. The inset shows a magnification of the
two-phonon contributions. }\\

\begin{figure}
\noindent
\includegraphics[width= 12cm]{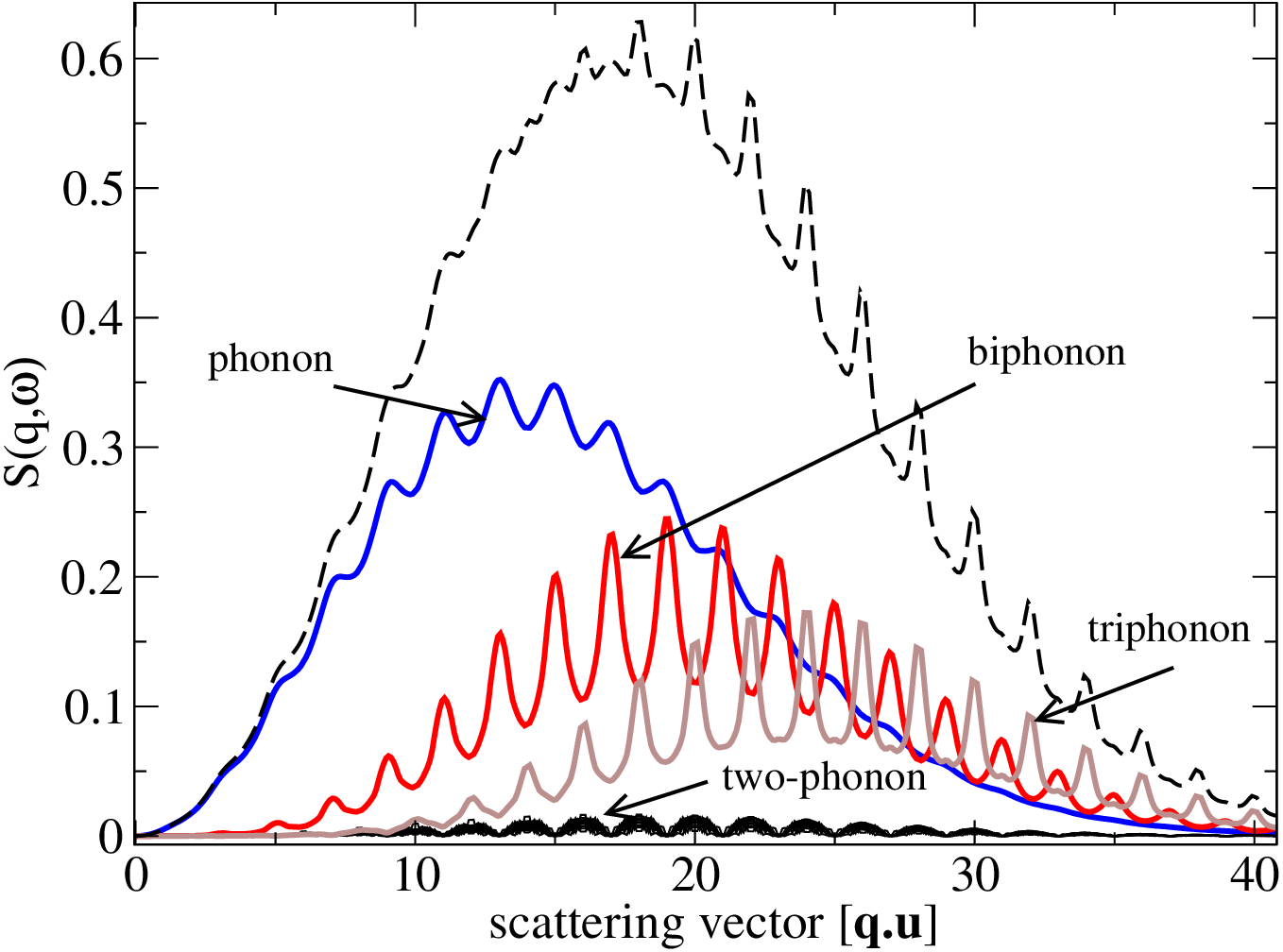}\\
\vspace{2cm}
\includegraphics[width= 12cm]{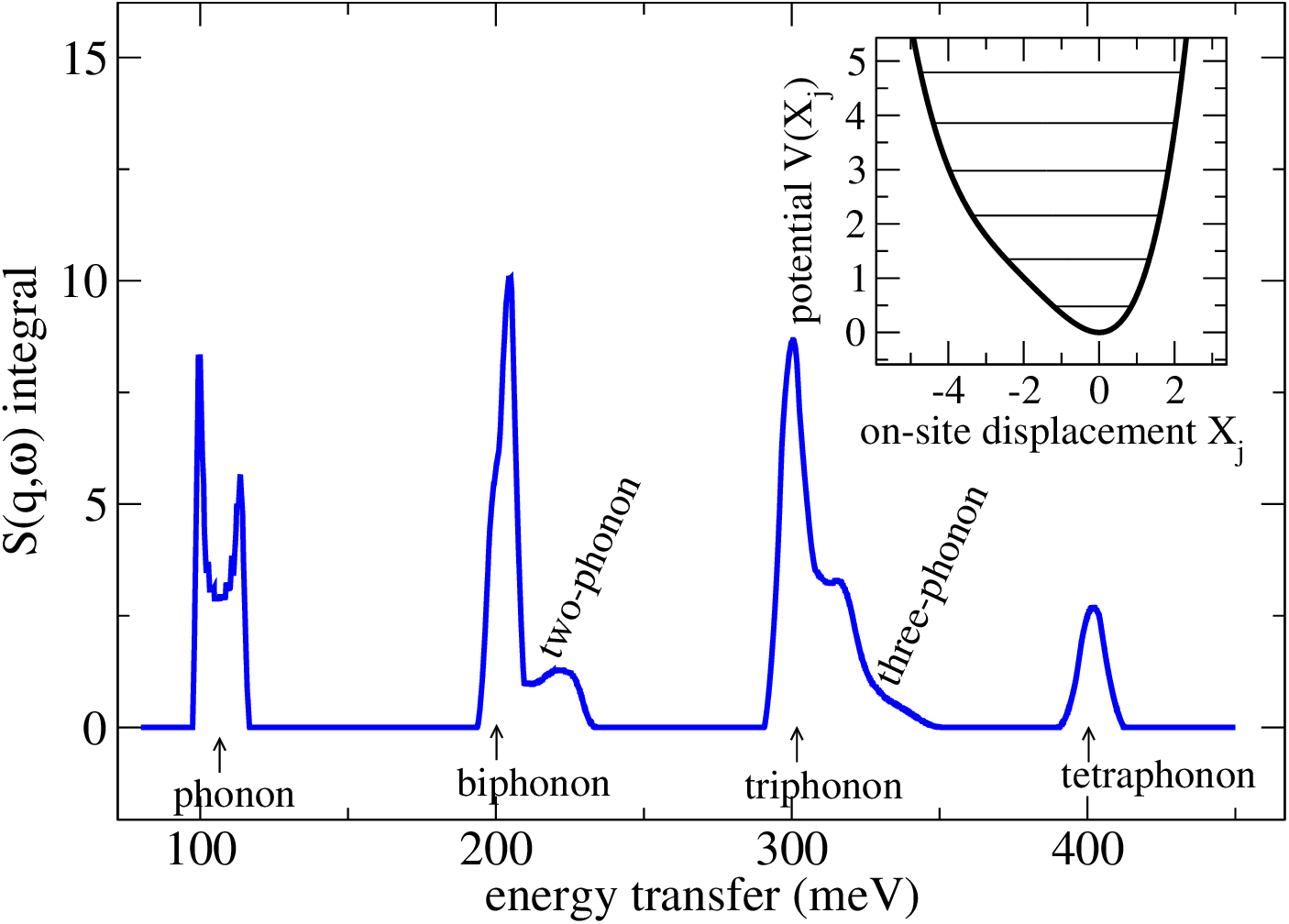}
\caption{\label{fig4} \bf{(2005) L. Proville}}
\begin{picture}(300,10)(0,0)
\put(-55,400){\makebox(0,0){\Large (a)}}
\put(-55,80){\makebox(0,0){\Large (b)}}
\end{picture}
\end{figure}

{Fig. \ref{fig4}:  In (a), the same as in Fig. \ref{fig3}(a) but
for different model parameters: $C=0.06$, $A_3=0.17$ and
$A_4=0.0231$. In addition, have been plotted the $S(q,\omega)$
profile for the triphonon and the integral of $S(q,\omega)$ over
the energy transfer $\omega$ (dashed line). The elastic
contribution has been skipped. In (b), a plot of the profile of
the $S(q,\omega)$ integral over the reciprocal lattice (see the
text), versus the energy transfer $\omega$, for same parameters as
in (a).  The inset in (b) shows the corresponding potential
landscape $V(X_j)$ and its quantum levels. Our energy unit
$\hbar\Omega$
has been fixed to $107$ meV.}\\

\newpage

\newpage

\newpage

\newpage
\newpage


\newpage


\begin{thebibliography}{99}
\bibitem{comment}
in a harmonic lattice, the exact formula for $S(q,\omega)$
is given in the text Sec.\ref{Sec3} and can be applied to the
the case of a linear KG
lattice (strictly quadratic) with a large intersite coupling ($C=1/2$).
\end{thebibliography}
\end{document}